\documentclass[letterpaper]{article} 
\PassOptionsToPackage{table}{xcolor}
\usepackage{aaai2026}  
\usepackage{times}  
\usepackage{helvet}  
\usepackage{courier}  
\usepackage[hyphens]{url}  
\usepackage{graphicx} 
\usepackage{natbib}  
\usepackage{caption} 
\usepackage{multirow}
\usepackage{array}
\usepackage{makecell}
\usepackage{subcaption}
\usepackage{booktabs}
\usepackage[many]{tcolorbox}
\usepackage{amssymb} 
\usepackage{algorithm}
\usepackage{algorithmic}
\usepackage{enumitem}
\usepackage{newfloat}
\usepackage{listings}
\DeclareCaptionStyle{ruled}{labelfont=normalfont,labelsep=colon,strut=off} 
\floatstyle{ruled}
\newfloat{listing}{tb}{lst}{}
\floatname{listing}{Listing}

\newcommand{\method}[0]{VerilogLAVD}
\definecolor{mygreen}{RGB}{219,239,197}
\nocopyright

\title{\method: LLM-Aided Rule Generation for Vulnerability Detection in Verilog}
\author{
    Xiang Long\textsuperscript{\rm 1},
    Yingjie Xia\textsuperscript{\rm 1}\thanks{Corresponding author.},
    Xiyuan Chen\textsuperscript{\rm 1},
    Li Kuang\textsuperscript{\rm 2}
}
\affiliations{
    \textsuperscript{\rm 1}Micro-Electronics Research Institute, Hangzhou Dianzi University\\
    \textsuperscript{\rm 2}School of Computer Science and Engineering, Central South University
}

\usepackage{bibentry}
\begin{document}
\maketitle
\begin{abstract}
Timely detection of hardware vulnerabilities during the early design stage is critical for reducing remediation costs. Existing early detection techniques often require specialized security expertise, limiting their usability. Recent efforts have explored the use of large language models (LLMs) for Verilog vulnerability detection. However, LLMs struggle to capture the structure in Verilog code, resulting in inconsistent detection results. To this end, we propose \method, the first LLM-aided graph traversal rule generation approach for Verilog vulnerability detection. Our approach introduces the Verilog Property Graph (VeriPG), a unified representation of Verilog code. It combines syntactic features extracted from the abstract syntax tree (AST) with semantic information derived from control flow and data dependency graphs. We leverage LLMs to generate VeriPG-based detection rules from Common Weakness Enumeration (CWE) descriptions. These rules guide the rule executor that traversal VeriPG for potential vulnerabilities. To evaluate \method, we build a dataset collected from open-source repositories and synthesized data. In our empirical evaluation on 77 Verilog designs encompassing 12 CWE types, \method~achieves an F1-score of 0.54. Compared to the LLM-only and LLM with external knowledge baselines, \method~improves F1-score by 0.31 and 0.27, respectively.
\footnote{Code will be released upon paper acceptance.}
\end{abstract}

\section{Introduction}

Modern hardware architectures have grown significantly more complex. 
This increased design complexity complicates the application of traditional security verification techniques, such as formal verification and dynamic simulation, which often suffer from scalability limitations and incomplete state-space coverage \cite{10159361}.
As a result, security vulnerabilities may remain undetected until the post-silicon validation stage, which substantially increases the cost and complexity of remediation \cite{karri2010trustworthy}. 
Therefore, ensuring compliance with security specifications during early design stages is essential to building trustworthy hardware systems.

Existing tools for analyzing security at the RTL have major limitations. The verification process requires engineers to manually review the RTL code to define security rules, create matching SystemVerilog Assertion (SVA) and run tests during simulation \cite{orenes2021autosva}. Engineers analyze the SVA results to identify potential security vulnerabilities. It only functions with fully completed hardware designs, making it unsuitable for early development stages. The RTL designs analysis tools in the early stage of hardware design, like Spyglass \cite{spyglass}, they can identify vulnerabilities early by using expert-defined static analysis rules. This workflow requires substantial security expertise and intensive labor for code review. More critically, Current these tools mainly focus on checking functional correctness instead of security.

\begin{figure}[!t]
    \centering
    \includegraphics[width=\linewidth]{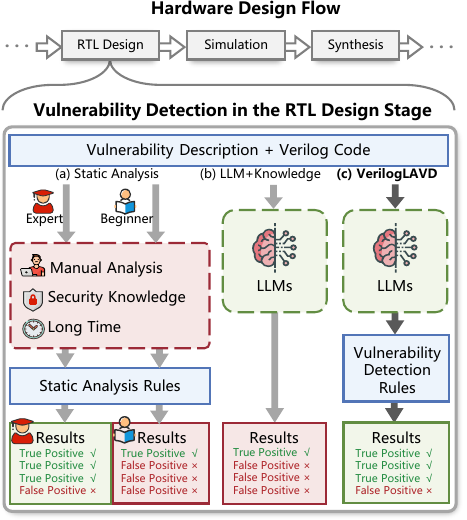}
    \caption{(a) Static analysis requires substantial time and expertise, and the effectiveness depends on the expertise and experience of the analysts. (b) LLM+Knowledge tends to generate false positives. (c) \method~reduces false positives by generating detection rules.}
    \label{fig:motivation}
\vspace{-5pt}
\end{figure}

\begin{figure*}[t]
\vspace{-5pt}
\centering
\includegraphics[width=2.1\columnwidth]{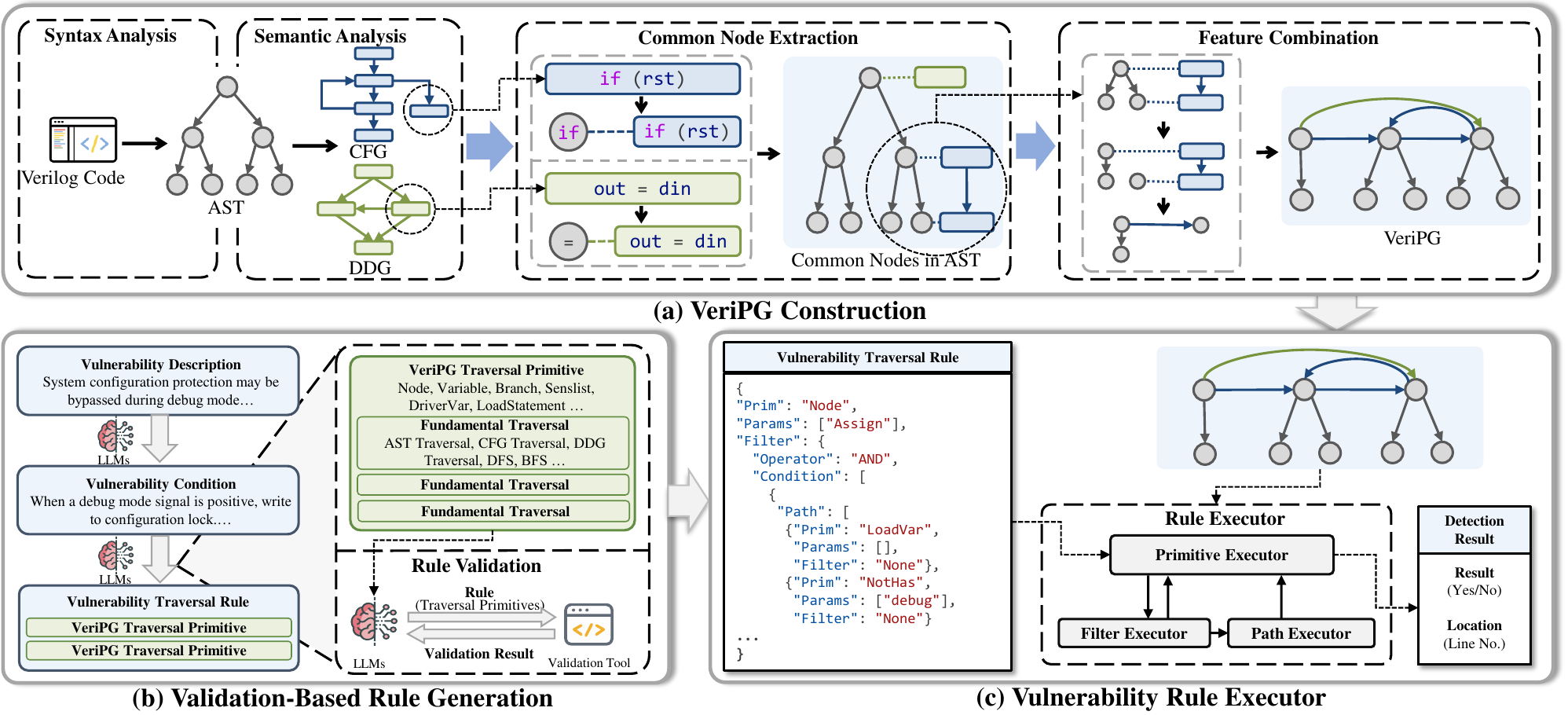}
\caption{Overview of \method. \method~consists of (a) \textit{VeriPG Construction}, (b) \textit{Validation-Based Rule Generation} and (c) \textit{Vulnerability Rule Execute}.}
\label{fig:overview}
\vspace{-5pt}
\end{figure*}

Recently, large language models (LLMs) have become widely used across many industries because LLMs are good at understanding context and logical reasoning, especially in software development domains, like Github Copilot \cite{githubcopilot} and Cursor \cite{cursor}. LLM is also applied in hardware security vulnerability detection, researchers use LLMs to find security weaknesses. Common methods create prompt templates that include both security knowledge and code structure information \cite{fang2025lintllm,saha2024empowering}. These prompt templates help LLMs detect vulnerabilities in the early stage of hardware design. For LLM, there is an important problem in vulnerability detection. LLMs are trained to excel at logical reasoning in natural language, they have a weak understanding of Verilog code structure. 
This issue makes LLMs much more likely to output incorrect information, producing unreliable results with factual errors.

Figure \ref{fig:motivation} illustrates the motivation of \method. As shown in Figure \ref{fig:motivation}(a), traditional approaches rely on manually crafted static analysis rules, which require extensive security expertise and significant time. Moreover, the effectiveness of these methods is highly dependent on the engineer's professional knowledge. In contrast, methods based on LLMs combined with external knowledge perform vulnerability detection through carefully designed prompts, as seen in Figure \ref{fig:motivation}(b). However, these methods often suffer from a high rate of false positives. This is primarily due to LLMs' limited understanding of Verilog’s structural semantics and their are misled by irrelevant information. This issue can be addressed by incorporating code graph structure analysis via static techniques.

To address the above challenges, we propose \method, a novel LLM-aided graph traversal rule generation approach for Verilog vulnerability detection. Inspired by Code Property Graph (CPG) \cite{yamaguchi2014modeling}, we aim to model Verilog code in a similar way. However, due to the syntactic and concurrency differences between Verilog and the languages targeted by CPG, we construct VeriPG, a graph-based representation of Verilog, and design a rule executor for vulnerability detection. 
We also design a vulnerability detection rule generation flow. Within this approach, LLMs are responsible for generating VeriPG traversal-based detection rules from natural language descriptions of vulnerabilities, which avoids direct structural reasoning. 
However, as LLMs tend to cause API misuse in software code generation, they similarly exhibit misuse of traversal primitives during rule generation \cite{zhong2024can}. We built a rule-validation tool to reduce the misuse of traversal primitive output by LLM. This tool monitors generated rules in real-time and gives instant feedback to LLMs to fix invalid rules. This tool reduces mistakes in vulnerability rules.

Our key contributions are as follows.

\begin{itemize}[leftmargin=*, itemsep=0.05mm]
    \item \textbf{VeriPG:} We propose VeriPG, a unified Verilog representation integrating AST, CFG, and DDG to enhance vulnerability detection accuracy and extensibility in detecting various vulnerability types.
    \item \textbf{Validation-Based Rule Generation:} We propose an improved generation method incorporating rule validation to verify rules, reducing rule misuse while improving correctness and stability.
    \item \textbf{Vulnerability Rule Executor:} We develop a suit of optimized traversal approach with a dedicated rule executor, leveraging Verilog characteristics and vulnerability patterns to significantly improve efficiency.
\end{itemize}

\section{Preliminaries}

\subsection{The Verilog Language}
Verilog is a hardware description language (HDL) most commonly employed at the register transfer level (RTL), which describes the flow of data between registers and the logical operations performed on those data. In this paper, all Verilog code under discussion is situated at the RTL. 
Unlike software programming languages, Verilog describes the structure and behavior of hardware components that operate in parallel. However, due to the concurrency and timing sensitivity of hardware, static analysis and vulnerability detection in Verilog pose unique challenges distinct from traditional software security analysis.

\subsection{Symbolic Definitions in VeriPG}
Formally, a Verilog module is represented as a directed graph $G=(V,E,A^V,A^E)$, where $V$ is AST node ($V_A$), $E$ contains $E_A$, $E_C$ and $E_D$, $A^V$ is property of $V$, contains $name$, $type$, $lineno$ and $value$, $A^E$ is property of $E$, contain $type$, $condition$.
The traversal primitives are specialized traversal functions for VeriPG. The symbolic definition of traversal primitives is as follows: $P(m)=\{n|V_{property}^A(n)=t_1,V_{property}^E(m,n)=t_2\}$, where $m$ is the current node, $n$ is the next node, $t_1$ nominates the type of $n$, $t_2$ nominates the type of edge between $m$ and $n$.
The graph construction is designed to preserve the structural, control flow, and data dependencies semantics of Verilog code, enabling precise traversal and reasoning for vulnerability detection. We will use this graph representation throughout the paper to define traversal primitives and vulnerability rules for CWE vulnerability detection.

\section{Methodology}
\subsection{Overview}
Figure~\ref{fig:overview} shows an overview of VerilogLAVD.
We begin by analyzing the Verilog code to construct its corresponding VeriPG representation. Simultaneously, we step-by-step extract vulnerability rules from natural language descriptions of CWE vulnerabilities by using LLM. These generated rules be validated to ensure their correctness. Subsequently, the generated VeriPG representation and the validated vulnerability rules are input into the vulnerability rules executor. This executor systematically examines the VeriPG structure against the vulnerability rules to identify potential security vulnerabilities.

\subsection{VeriPG Construction}
The VeriPG construction phase consists of three key steps: 1) \textit{Syntax and Semantic Analysis}, 2) \textit{Extraction of Common Nodes}, and 3) \textit{Feature Combination}. 
We first analyzes Verilog code to generate three graph representations: AST, CFG and DDG.
Subsequently, we identifies the common nodes in these structures.
Finally, we merges the three graphs into a unified model by leveraging these common nodes.

\subsubsection{Syntax and Semantic Analysis} We employ the PyVerilog toolkit to parse Verilog code and generate its AST. Custom scripts then perform semantic analysis on the AST to extract both CFG and DDG. We construct CFG in procedural block of Verilog. The CFG construction involves identifying procedural blocks (e.g., \texttt{Always} blocks) and control blocks (such as \texttt{IfStatement} and \texttt{ForStatement}), followed by establishing control flow edges between them. Unlike the CFG of software program language, the CFG of Verilog requires handle parallel statements. 
For DDG generation, we analyze signal definitions, usages, dependency relationships, and construct corresponding dependency edges. This process yields three complementary representations of the Verilog code: AST, CFG and DDG.

\subsubsection{Common Node Extraction} Our analysis of these structures begins with extracting appropriate syntactic identifiers from the AST to serve as common nodes for representation fusion. While CFG nodes ($V_C$) and DDG nodes ($V_D$) represent complete code statements, each containing distinctive identifiers or their combinations that indicate semantic types ($\exists v_{common} \in  V_A, \exists v_C\in V_C,\exists v_D \in V_D, v_{common} \in v_C ~\text{or}~ v_{common} \in v_D$), these identifiers maintain correspondence with their AST counterparts. We systematically extract these aligned identifiers as fusion anchors for subsequent feature integration ($V_{common} \cup V_A \to V$).

\subsubsection{Feature Combination} The feature fusion process uses the AST as the foundational structure, augmented with control flow edges from the CFG and data dependency edges from the DDG. To prepare for fusion, we preprocess the AST by breaking the AST edges between common nodes ($e_A\in E_A,  e_A=(v_{common}^i,v_{common}^j) \to e_A=\varnothing$) and decomposing it into multiple syntax subtrees rooted at the identified common nodes, each subtree representing a code segment that corresponds to the CFG / DDG nodes. Semantic enrichment is achieved by incorporating the control edges of CFG and the dependency edges of DDG between the corresponding subtree nodes ($\exists v_{common}^i\in v_C^i,\exists v_{common}^j\in v_C^j,e_C=(v_C^i,v_C^j)\to e_C=(v_{common}^i,v_{common}^j)\ \text{similarly }e_D=(v_{common}^m,v_{common}^n$). This integrated representation, termed VeriPG, effectively combines the three structural code representations through their interconnected common nodes.

\subsection{Validation-Based Rule Generation}
To mitigate reliance on security experts in vulnerability analysis, we leverage LLMs to generate vulnerability rules through natural language descriptions of vulnerabilities. However, the inherent output randomness of LLMs introduces instability in rule generation. To address this, we propose a validation-based vulnerability rule generation methodology.
We first extract the vulnerability conditions from the CWE descriptions by LLMs. 
Subsequently, we generate the VeriPG traversal rules for vulnerability detection using the rule validation tool.

\begin{figure}[t]
\centering
\includegraphics[width=1\columnwidth]{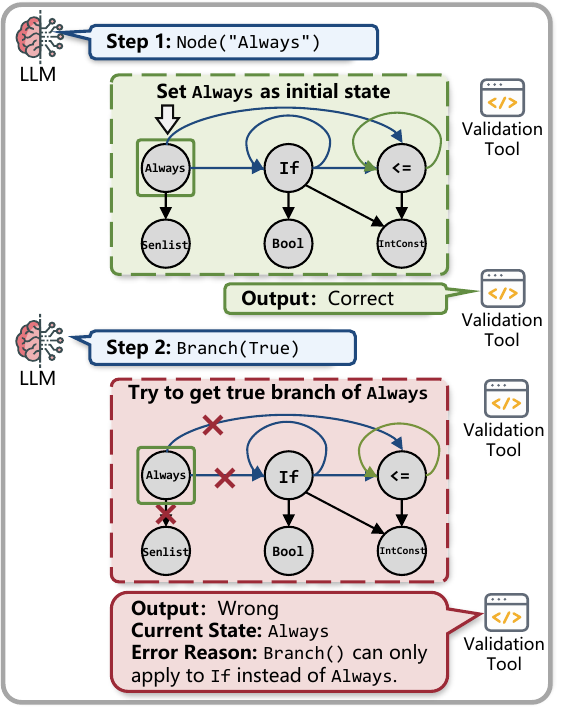}
\caption{An illustration of Rule Validation process. LLM (Left) generates traversal primitive step by step, Validation Tool (Right) validates traversal primitive and output result.}
\label{fig:path-validation}
\end{figure}

\subsubsection{VeriPG Traversal Primitive} We propose a set of fundamental traversal methods for VeriPG that synergistically consider Verilog syntax characteristics and common vulnerability patterns, achieving enhanced efficiency in vulnerability detection. These basic traversal methods are systematically combined through textual vulnerability descriptions, forming configurable vulnerability rules. The traversal primitives consist of three categories: 1) \textit{Generic graph traversal}, contains fundamental traversals, like DFS, BFS, traversal along AST edge; These traversals are basic components of other traversal primitives. 2) \textit{Boolean operations}, mainly involve the composition and computation of the results from other traversal primitives. 3) \textit{VeriPG-specific traversal}, focuses on semantic-aware processing of Verilog syntax elements through direct indexing of critical syntax nodes and specialized traversal strategies. For instance, our conditional variable traversal for \texttt{IfStatement} nodes identifies variables affecting branch decisions, while rule analysis leverages CFG edges to determine execution rules under specific conditions. The assignment variable traversal utilizes DDG edges to track variable propagation. Generic graph traversal supplements these specialized methods with conventional graph operations including node type filtering and depth-first search along specific edge types. Boolean operations enable logical combinations of traversal results, supporting complex pattern detection through result transformation and filtering. Examples of two primitives (\texttt{Node()} and \texttt{Branch()}) are illustrated in Equations (\ref{eq:node}) and (\ref{eq:branch}).
\begin{equation}  
\label{eq:node}
\text{Node}(t)=\{n|V^A_{type}(n)=t \}
\end{equation}
where $t$ is one kind of node $type$, $V^A_{type}(n)$ means obtaining $type$ property value of $n$, this primitive returns all nodes with type $t$.
\begin{align}
\label{eq:branch}
\text{Branch}(m)  = \{ n \mid & V^A_{\text{type}}(m)=\text{IfStatement} \& \notag \\
& V^E_{\text{type}}(m,n)=\text{CFG} \}
\end{align}
where \texttt{IfStatement} is value of node $type$, CFG is value of edge $type$, $V^E_{type}(m,n)$ means get $type$ property value of edge from $m$ to $n$, this primitive returns all nodes reachable from an \texttt{IfStatement} node along CFG edges.
\subsubsection{Vulnerability Detection Rule} Vulnerability rules implement combinatorial logic through three core components: Functions, Filters, and Paths. Functions encapsulate basic traversal operations with configurable parameters and result filters. Filters apply Boolean logic (AND/OR) to combine multiple verification conditions, which can be either nested Functions or Paths. Paths define sequential execution chains where preceding Function outputs serve as subsequent Function inputs. For practical implementation, these rules are stored in JSON format with structured schema definitions.

\subsubsection{Vulnerability Condition Extraction} Our solution try to bridge the significant semantic gap between natural language descriptions and formal vulnerability rules through a Chain-of-Thought (CoT) approach. CWE descriptions typically contain two critical components: vulnerability manifestations and root causes, which essentially define vulnerability types through specific conditional constraints. Our method systematically extracts these constraint from textual descriptions and converts them into vulnerability judgment conditions. Through iterative optimization of prompt engineering, we achieve stable and standardized LLM outputs.

\subsubsection{Rule Generation}
To address two key issues—(1) hallucinated rules that exceed VeriPG’s structural constraints, and (2) invalid traversal primitives generated during rule synthesis—we implement a rule validation mechanism. This mechanism incorporates a VeriPG-based node state machine that continuously monitors transformation states throughout the rule generation process. It provides real-time feedback on node state transitions and validates transformation steps, enabling the LLM to dynamically correct erroneous outputs. The resulting validated vulnerability rules are directly applicable to the automated vulnerability rule executor.

\begin{table}[!b]
\centering
\caption{Vulnerability categories (this work), their corresponding CWE types and number of designs in the dataset.}
\label{tab:VulCategory}
\renewcommand{\arraystretch}{1}
\small
\begin{tabular}{l c c}
\toprule
\textbf{Category} & \textbf{CWEs} & \textbf{No.} \\ 
\midrule
Improper Access Control & 1231, 1243, 1244, 1280 & 17 \\
Improper Resource Operate & 226, 1258, 1271 & 13 \\
Improper Lock & 1232, 1234 & 11 \\
Side Channel & 1255, 1300 & 11 \\
Finite State Machine & 1245 & 7 \\
Non-Vulnerability & None & 12 \\
\bottomrule
\end{tabular}
\vspace{-5pt}
\end{table}

\begin{table*}[!t]
\centering
\caption{Vulnerability detection performance evaluation of recent LLMs and \method. F1 scores are shown in bold, and the best-performing metrics are highlighted with a green background.}
\label{tab:experiment}
\setlength{\tabcolsep}{3pt} 
\renewcommand{\arraystretch}{1.3}
\resizebox{\textwidth}{!}{ 
\large 
\begin{tabular}{
l||c c c|c c c|c c c||c c c|c c c|c c c
}
\hline
\multirow{2}{*}{\textbf{Vulnerability Category}} 
    & \multicolumn{3}{c|}{\makecell{\textbf{DeepSeek-V3}}} 
    & \multicolumn{3}{c|}{\makecell{\textbf{DeepSeek-V3}\\\textbf{+Knowledge}}} 
    & \multicolumn{3}{c||}{\makecell{\textbf{DeepSeek-V3}\\\textbf{+\method}}} 
    & \multicolumn{3}{c|}{\textbf{GPT-4o}} 
    & \multicolumn{3}{c|}{\makecell{\textbf{GPT-4o}\\\textbf{+Knowledge}}} 
    & \multicolumn{3}{c}{\makecell{\textbf{GPT-4o}\\\textbf{+\method}}} \\
\cline{2-4} \cline{5-7} \cline{8-10} \cline{11-13} \cline{14-16} \cline{17-19}
    & P(\%) & R(\%) & F1 & P(\%) & R(\%) & F1 & P(\%) & R(\%) & F1 & P(\%) & R(\%) & F1 & P(\%) & R(\%) & F1 & P(\%) & R(\%) & F1 \\
\hline
\hline
\textbf{Improper Access Control} & 15.07 & 64.71 & \textbf{0.24} & 25.71 & 52.94 & \textbf{0.35} & \cellcolor{mygreen}48.15 & \cellcolor{mygreen}76.47 & \cellcolor{mygreen}\textbf{0.59} & 11.71 & 76.47 & \textbf{0.20} & 14.85 & \cellcolor{mygreen}88.24 & \textbf{0.25} & \cellcolor{mygreen}60.87 & 82.35 & \cellcolor{mygreen}\textbf{0.70} \\
\textbf{Improper Resource Operate} & 13.79 & 30.77 & \textbf{0.19} & \cellcolor{mygreen}26.09 & 46.15 & \textbf{0.33} & 24.32 & \cellcolor{mygreen}69.23 & \cellcolor{mygreen}\textbf{0.36} & 12.07 & 53.85 & \textbf{0.20} & 12.90 & 61.54 & \textbf{0.21} & \cellcolor{mygreen}32.26 & \cellcolor{mygreen}76.92 & \cellcolor{mygreen}\textbf{0.45} \\
\textbf{Improper Lock} & 31.25 & 45.45 & \textbf{0.37} & 35.29 & 54.55 & \textbf{0.43} & \cellcolor{mygreen}43.75 & \cellcolor{mygreen}63.64 & \cellcolor{mygreen}\textbf{0.52} & 24.24 & \cellcolor{mygreen}72.73 & \textbf{0.36} & 21.88 & 63.64 & \textbf{0.33} & \cellcolor{mygreen}46.15 & 54.55 & \cellcolor{mygreen}\textbf{0.50} \\
\textbf{Side Channel} & 18.75 & 27.27 & \textbf{0.22} & 26.67 & 36.36 & \textbf{0.31} & \cellcolor{mygreen}83.88 & \cellcolor{mygreen}45.45 & \cellcolor{mygreen}\textbf{0.59} & 8.16 & 36.36 & \textbf{0.13} & 12.50 & \cellcolor{mygreen}72.73 & \textbf{0.21} & \cellcolor{mygreen}53.85 & 63.64 & \cellcolor{mygreen}\textbf{0.58} \\
\textbf{Finite State Machine} & 11.11 & 14.29 & \textbf{0.13} & 17.65 & 42.86 & \textbf{0.25} & \cellcolor{mygreen}45.45 & 
\cellcolor{mygreen}71.43 & \cellcolor{mygreen}\textbf{0.56} & 18.75 & 42.86 & \textbf{0.26} & 27.27 & 42.86 & \textbf{0.33} & \cellcolor{mygreen}54.55 & \cellcolor{mygreen}85.71 & \cellcolor{mygreen}\textbf{0.67} \\
\hline
\textbf{Total} & 16.78 & 40.68 & \textbf{0.24} & 26.17 & 47.46 & \textbf{0.34} & \cellcolor{mygreen}40.21 & \cellcolor{mygreen}66.10 & \cellcolor{mygreen}\textbf{0.50} & 13.11 & 59.32 & \textbf{0.21} & 15.19 & 69.49 & \textbf{0.25} & \cellcolor{mygreen}47.25 & \cellcolor{mygreen}72.88 & \cellcolor{mygreen}\textbf{0.57} \\
\hline
\end{tabular}
}
\end{table*}

\subsubsection{Rule Validation} The rule validation tool uses a finite state machine built from the VeriPG's structure. We treat different VeriPG nodes as states and connections between nodes as state transitions. Some nodes have self-loop edges that show nested relationships, such as when one \texttt{If} statement contains another. The tool sets a start state and moves between states by traversal primitives. The process of rule validation begins at a common node decided by LLM, and state transitions are driven by traversal primitives. When multiple valid transitions exist for a traversal primitive, the tool tracks all possible state transition branches, but illegal branches in successor state transition will be stopped tracking immediately. This method checks whether the LLM's traversal commands follow VeriPG's rules and provides real-time feedback, as shown in Figure \ref{fig:path-validation}.

\subsection{Vulnerability Rule Executor}
We design a rule executor to traverse the VeriPG by parsing vulnerability rules. By jointly optimizing traversal primitives and adopting an extensible rule architecture, the executor achieves both high traversal efficiency and flexible support for new vulnerability types. The rule executor comprises three key components: the \textit{Primitive Executor}, which serves as the core and entry point responsible for executing traversal primitives of rules on VeriPG; It first resolves its arguments (primitives or paths) by invoking the corresponding executor, then performs the traversal and applies a filter to extract results that meet the specified conditions; the \textit{Filter Executor}, which applies filtering conditions during rule execution; and the \textit{Path Executor}, which manages and interprets the traversal paths specified by the rules.

\section{Experiments}
\subsection{Experiments Setup}

We evaluate \method~in different aspects to answer the following research questions.
\textit{RQ1: How effective is \method~compared with LLM-only approaches across different categories of vulnerabilities?}
\textit{RQ2: How effective is Rule Validation of vulnerability detection rule generation?}
\textit{RQ3: How efficient is the vulnerability rule executor across Verilog designs of different scales?}
We introduce a representative case to prove the effectiveness of \method.

\subsubsection{Dataset}
Existing hardware CWE datasets cover only a limited range of vulnerability types, making it difficult to comprehensively evaluate our approach. To address this limitation, we constructed a customized dataset based on CWE classifications by using samples from the real-world open-source repository Hack@21 as seed data, while also accounting for the diverse code structures that can occur within the same CWE category. Our mutation strategies include: 1) \textit{Name substitution}, where signal or register names related to the vulnerability are replaced; 2) \textit{Process block extension}, where unrelated procedural blocks are inserted to increase code complexity and length; and 3) \textit{Structural complication}, which involves extending the distance between vulnerable components within the source file or introducing additional branching and looping constructs. Using these methods, we generated 59 positive samples. We obtained 18 negative samples from the corresponding patched versions. The final dataset comprises 77 Verilog designs covering 12 CWE types, as shown in Table \ref{tab:VulCategory}.

\subsubsection{Baseline}
Although several related studies and tools have achieved promising results on this task, we did not adopt them as direct baselines for the following reasons: 1) lack of publicly available code, 2) inconsistency in research objectives, and 3) heavy reliance on human expertise. Instead, we compare our approach against the LLM-only and LLM+Knowledge methods. 
In the LLM-only method, we provide the LLMs with the Verilog code along with a brief description of the corresponding CWE vulnerability. The LLM+Knowledge method extends this input by incorporating detailed CWE information retrieved from the official MITRE CWE database.

In all methods, we using two representative LLMs: GPT-4o \cite{openai2024gpt4o} and DeepSeek-V3 \cite{deepseekai2024deepseekv3technicalreport}. GPT-4o is widely recognized for its strong performance and has served as a benchmark for many subsequent models since its release. DeepSeek-V3, by contrast, is an open-source model that offers a favorable balance between accuracy and computational efficiency, making it a practical option for a broad range of real-world applications.

\subsubsection{Implementation Detail}
We implemented the whole approach using Python 3.10, parsed the Verilog code into an AST using PyVerilog \cite{takamaeda2015pyverilog}, used neo4j \cite{10.1145/2384716.2384777} to operate the VeriPG graph structure, and built the agent in the Vulnerability Rule generation Module using Autogen \cite{wu2024autogen}.

\subsubsection{Evaluation Metrics}
We choose precision (P), recall (R), and F1 score as our evaluation metrics, which are widely used in previous studies. For each individual CWE, the dataset contains an imbalanced number of positive and negative samples. Therefore, we avoid using accuracy as an evaluation metric, as it can be misleading under such conditions.

\subsection{Effectiveness of \method~(RQ1)}
To evaluate the effectiveness of \method~in RTL vulnerability detection, we conducted a series of comparative experiments against representative baselines across different CWE vulnerability categories.
We adopted the Pass@5 strategy to mitigate the randomness of LLM outputs. Each method was executed independently 5 times, and a Verilog sample was considered vulnerable if any of the 5 runs detected a vulnerability. This evaluation covered 12 types of CWE-defined vulnerabilities, which we grouped into 5 broader categories to facilitate analysis.

Table~\ref{tab:experiment} summarizes the performance comparison between \method~and the LLM+Knowledge baseline using different backbone models. The results indicate that \method~consistently outperforms the knowledge-augmented approach, demonstrating greater robustness across diverse model architectures.
For instance, DeepSeek-V3+\method~achieved a 44.12\% improvement in F1 score over DeepSeek-V3+Knowledge, while GPT-4o+\method~showed a 133.33\% increase compared to GPT-4o+Knowledge.
These findings confirm the effectiveness of \method~in accurately identifying RTL security vulnerabilities and highlight its potential as a generalizable framework that surpasses conventional LLM fine-tuning and prompt engineering methods.

\subsection{Impact of Rule Validation (RQ2)}
To assess the impact of rule validation on vulnerability rule generation, particularly in mitigating traversal primitive misuse, we conduct a series of ablation experiments using the DeepSeek-V3 backbone model.
We compare three configurations: 1) \textit{LLM+Knowledge (DeepSeek-V3)}, 2) \textit{Rule Extraction without Rule Validation}, and 3) \textit{Rule Generation with Rule Validation}.

\begin{table}[t]
\centering
\small
\caption{
The results of ablation study.
}
\label{tab:centered_multiline}
\renewcommand{\arraystretch}{1.3}
\setlength{\tabcolsep}{4pt}
\begin{tabular}{l c c c}
\toprule
\multirow{2}{*}{\makecell[c]{\textbf{Method}}} & \multicolumn{3}{c}{\textbf{Traversal Primitive Misuse}} \\
\cmidrule{2-4}
 & IPTR (\%) & IPMR (\%) & Total (\%) \\
\midrule
\textbf{LLM+Knowledge} & 27.04 & 13.83 & 40.87 \\
\textbf{\method~(w/o PV)} & 23.96 & 12.88 & 36.84 \\
\textbf{\method~(with PV)} & \textbf{3.49} & \textbf{6.61} & \textbf{10.10} \\
\bottomrule
\end{tabular}
\end{table}

\begin{figure}[t]
    \centering
    \includegraphics[width=\linewidth]{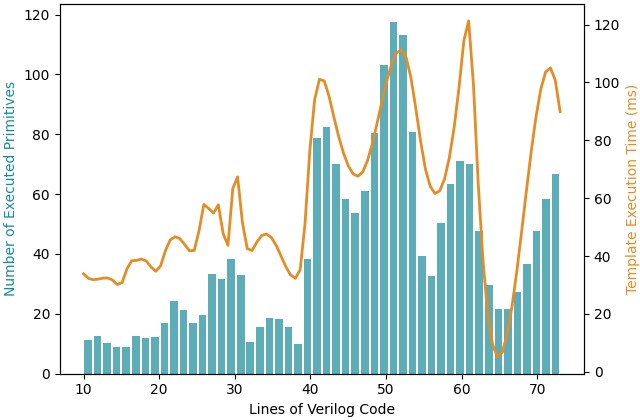}
    \caption{Variation of number of executed primitives and rule execution time with the lines of Verilog code.}
    \label{fig:RQ3}
\end{figure}

All configurations use the same inputs: CWE descriptions, VeriPG structural information, traversal primitives, and example rule. In the first configuration, the LLM directly receives these inputs as a prompt. The second converts CWE descriptions into rules through step-by-step reasoning. The third further refines these rules by applying a rule validation tool iteratively, halting upon success or after 50 iterations.
Each method runs five times per CWE type across all 12 categories. The resulting rules are manually examined for two types of traversal primitive misuse: Illegal rules, where a primitive is incorrectly applied to an unrelated node, and Illegal parameters, involving invalid parameter counts or types.
We define two metrics: the illegal rule rate and illegal parameter rate, calculated as the proportion of violations among all generated primitives.

Table~\ref{tab:centered_multiline} presents the results. The rule validation configuration significantly lowers the overall misuse rate to 10.10\%, compared to 40.87\% for the baseline and 36.84\% for the no-validation method. Specifically, it reduces the illegal rule rate to 3.49\%, representing decreases of 87.10\% and 85.43\%, respectively. The illegal parameter rate also drops to 6.61\%, yielding reductions of 52.21\% and 48.68\% over the two other methods.
These findings demonstrate that rule validation substantially reduces misuse, particularly illegal traversal rules, thereby improving the structural integrity and reliability of generated vulnerability rules.

\begin{figure}[t]
    \centering
    \begin{subfigure}[b]{\linewidth}
        \includegraphics[width=\linewidth]{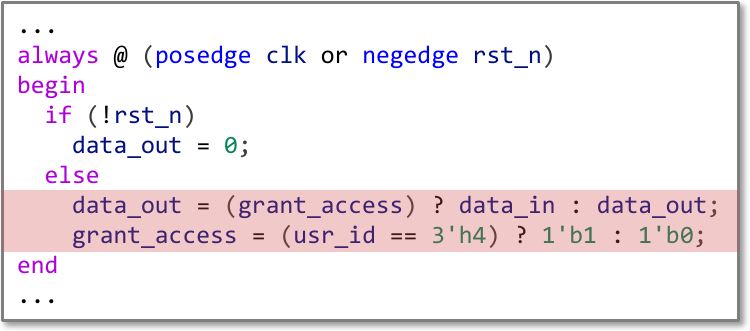}
        \caption{Verilog Code with Vulnerability}
        \label{fig:RQ4a}
    \end{subfigure}
    
    \vspace{\floatsep} 
    
    \begin{subfigure}[b]{\linewidth}
        \includegraphics[width=\linewidth]{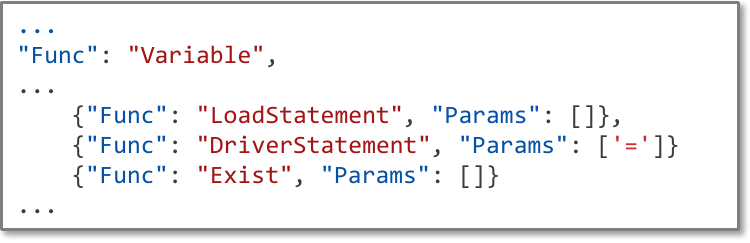}
        \caption{Vulnerability Rule}
        \label{fig:RQ4b}
    \end{subfigure}
    
    \caption{(a) A Verilog code with CWE-1280. The code containing  vulnerabilities is highlighted with a red background. (b) The fragment of CWE-1280 vulnerability rule.}
    \label{fig:RQ4}
\end{figure}

\subsection{Efficiency Across Verilog Design Scales (RQ3)}
To evaluate the performance of the vulnerability rule executor on single-module Verilog code of varying lengths, we conducted experiments using our dataset.
As shown in Figure \ref{fig:RQ3}, 
we observe that as the number of code lines increases, neither the number of executed traversal primitives nor the rule execution time exhibits a clear upward trend. This suggests that the executor’s performance is not directly correlated with code length. Notably, the number of traversal primitives closely aligns with execution time, indicating that primitive execution dominates the overall scanning process.
Interestingly, in some cases, longer Verilog files yield shorter execution times, suggesting that the scanner is largely insensitive to irrelevant code and can maintain high performance even on lengthy designs.

\subsection{Case Study}
We present a representative RTL vulnerability case in which data-flow analysis is critical for accurate detection. 
Specifically, detecting CWE-1280 requires analyzing data dependencies: this vulnerability arises when an access control condition is used before being properly initialized, potentially leading to nondeterministic execution behavior,as shown in Figure \ref{fig:RQ4a}. To efficiently detect such issues, our approach uses customized traversal primitives for the CFG and DDG.

As shown in Figure \ref{fig:RQ4b}, the corresponding vulnerability rule consists of four key steps:
1) Use the \texttt{Variable} primitive to collect all signals in the module, then apply a Filter to select relevant ones;
2) Traverse from each selected signal to its usage via \texttt{LoadStatement}, following data dependency edges;
3) Move from the usage point to the corresponding blocking assignment using \texttt{AssignStatement} along the DDG;
4) Apply the \texttt{Exist} primitive to check whether the identified pattern exists and return a boolean result.
This rule demonstrates how VeriPG utilizes data dependency information for graph-based traversal, emphasizing its essential role in enabling semantic-level vulnerability detection.
More details/results are referred to the appendix.

\section{Related Work}
\paragraph{Traditional Hardware Security Verification}
With the increasing complexity of hardware design, a variety of hardware security verification methods have emerged. 1) \textit{Simulation-based Validation} depends on manually crafted test cases to assess hardware security \cite{dessouky2019hardfails, rajendran2023hunter}; however, this method is labor-intensive, inefficient, and suffers from limited coverage. 2) \textit{Fuzzing Testing and Penetration Testing} employ automated test case generation via mutation and feedback mechanisms \cite{hossain2023socfuzzer,al2023sharpen}, but they require extensive testing to achieve sufficient coverage. 3) \textit{Assertion-based Formal Verification} expresses security properties using SVAs and checks them with formal tools \cite{aftabjahani2021special, orenes2021autosva}; despite its rigor, this approach often encounters state space explosion in complex designs and demands considerable expertise to construct accurate SVAs. 4) \textit{Security Analysis based on RTL Code Structure} leverages RTL code structure scanning with problem-specific detectors \cite{ahmad2022don}. While effective for certain issues, it requires custom analyzer development for different vulnerability types, heavily relying on domain-specific knowledge. 5) \textit{Information Flow Tracking (IFT)} monitors the propagation of sensitive data to identify unauthorized flows \cite{solt2022cellift,zhao2024static}, but it introduces significant computational overhead. 6) \textit{Machine Learning-based Verification}, including methods based on graph neural network (GNN), strives to automate the validation process \cite{fan2024efficient, yasaei2022hardware}; however, the absence of large-scale labeled datasets limits the generalization capability of these models.

\paragraph{LLM-aided Hardware Security Verification}
Recent advances have explored the application of LLMs in various aspects of hardware security verification. 1) \textit{LLM-based Logical Verification} utilizes LLMs to reason about and identify issues in RTL code logic \cite{akyash2024self,pearce2025asleep,zhang2024empirical}; however, the inherent complexity and structural characteristics of RTL present significant challenges for accurate comprehension by LLMs. 2) \textit{LLM-aided Formal Verification} employs LLMs to generate SVAs, aiming to lower the barrier to writing formal properties \cite{orenes2023using}. However, due to the intricate nature of SVAs and the ambiguity often found in security specifications, the generated results frequently exhibit instability. 3) \textit{LLM-aided IFT} improves the efficiency and scalability of IFT by dynamically inferring security properties \cite{mashnoor2025llm}. 

Despite these promising directions, prior work has not investigated the integration of LLMs with RTL-structure-based security analysis. Leveraging LLMs to generate analysis rules can reduce the manual effort involved in rule construction, thereby enhancing the scalability of RTL structural analysis and facilitating early-stage security assessment during hardware design.

\section{Conclusion and Future Work}
In this paper, we proposed the first LLM-aided graph traversal rule generation approach for Verilog vulnerability detection. Our method introduces VeriPG, a unified intermediate representation that integrates syntactic and semantic information from ASTs, CFGs, and DDGs. Using LLMs, we analyze vulnerability patterns and generate graph traversal rule for structured vulnerability matching. \method~achieving an F1 score improvement of 0.31 and 0.25 over LLM-only and LLM+knowledge baselines, respectively.

During the course of this study, we observed that Verilog vulnerabilities tend to be concentrated within specific functional modules. However, our current approach does not incorporate functional context into the detection process. In future work, we plan to explore integrating functional module information to improve vulnerability localization and detection accuracy.


\begin{thebibliography}{50}
\providecommand{\natexlab}[1]{#1}

\bibitem[{Aftabjahani et~al.(2021)Aftabjahani, Kastner, Tehranipoor, Farahmandi, Oberg, Nordstrom, Fern, and Althoff}]{aftabjahani2021special}
Aftabjahani, S.; Kastner, R.; Tehranipoor, M.; Farahmandi, F.; Oberg, J.; Nordstrom, A.; Fern, N.; and Althoff, A. 2021.
\newblock Special session: Cad for hardware security-automation is key to adoption of solutions.
\newblock In \emph{2021 IEEE 39th VLSI Test Symposium (VTS)}, 1--10. IEEE.

\bibitem[{Ahmad et~al.(2022)Ahmad, Liu, Collini, Pearce, Fung, Valamehr, Bidmeshki, Sapiecha, Brown, Chakrabarty et~al.}]{ahmad2022don}
Ahmad, B.; Liu, W.-K.; Collini, L.; Pearce, H.; Fung, J.~M.; Valamehr, J.; Bidmeshki, M.; Sapiecha, P.; Brown, S.; Chakrabarty, K.; et~al. 2022.
\newblock Don't CWEAT it: Toward CWE analysis techniques in early stages of hardware design.
\newblock In \emph{Proceedings of the 41st IEEE/ACM International Conference on Computer-Aided Design}, 1--9.

\bibitem[{Akter, Khalil, and Bayoumi(2023)}]{10159361}
Akter, S.; Khalil, K.; and Bayoumi, M. 2023.
\newblock A Survey on Hardware Security: Current Trends and Challenges.
\newblock \emph{IEEE Access}, 11: 77543--77565.

\bibitem[{Akyash and Kamali(2024)}]{akyash2024self}
Akyash, M.; and Kamali, H.~M. 2024.
\newblock Self-hwdebug: Automation of llm self-instructing for hardware security verification.
\newblock In \emph{2024 IEEE Computer Society Annual Symposium on VLSI (ISVLSI)}, 391--396. IEEE.

\bibitem[{Al-Shaikh et~al.(2023)Al-Shaikh, Vafaei, Rahman, Azar, Rahman, Farahmandi, and Tehranipoor}]{al2023sharpen}
Al-Shaikh, H.; Vafaei, A.; Rahman, M. M.~M.; Azar, K.~Z.; Rahman, F.; Farahmandi, F.; and Tehranipoor, M. 2023.
\newblock Sharpen: Soc security verification by hardware penetration test.
\newblock In \emph{Proceedings of the 28th Asia and South Pacific Design Automation Conference}, 579--584.

\bibitem[{Anysphere(2023)}]{cursor}
Anysphere. 2023.
\newblock The AI Code Editor.
\newblock \url{https://cursor.com/}.
\newblock Accessed: 2025-07-11.

\bibitem[{DeepSeek-AI(2024)}]{deepseekai2024deepseekv3technicalreport}
DeepSeek-AI. 2024.
\newblock DeepSeek-V3 Technical Report.
\newblock arXiv:2412.19437.

\bibitem[{Dessouky et~al.(2019)Dessouky, Gens, Haney, Persyn, Kanuparthi, Khattri, Fung, Sadeghi, and Rajendran}]{dessouky2019hardfails}
Dessouky, G.; Gens, D.; Haney, P.; Persyn, G.; Kanuparthi, A.; Khattri, H.; Fung, J.~M.; Sadeghi, A.-R.; and Rajendran, J. 2019.
\newblock $\{$HardFails$\}$: insights into $\{$software-exploitable$\}$ hardware bugs.
\newblock In \emph{28th USENIX Security Symposium (USENIX Security 19)}, 213--230.

\bibitem[{Fan et~al.(2024)Fan, Tang, Sun, Liu, and Li}]{fan2024efficient}
Fan, R.; Tang, Y.; Sun, H.; Liu, J.; and Li, H. 2024.
\newblock An efficient ml-based hardware trojan localization framework for rtl security analysis.
\newblock In \emph{Proceedings of the 2024 ACM/IEEE International Symposium on Machine Learning for CAD}, 1--7.

\bibitem[{Fang et~al.(2025)Fang, Chen, Yang, Guo, Dai, and Wang}]{fang2025lintllm}
Fang, Z.; Chen, R.; Yang, Z.; Guo, Y.; Dai, H.; and Wang, L. 2025.
\newblock Lintllm: An open-source verilog linting framework based on large language models.
\newblock \emph{arXiv preprint arXiv:2502.10815}.

\bibitem[{Github(2021)}]{githubcopilot}
Github. 2021.
\newblock Introducing GitHub Copilot: your AI pair programmer.
\newblock \url{https://github.blog/news-insights/product-news/introducing-github-copilot-ai-pair-programmer/}.
\newblock Accessed: 2025-07-11.

\bibitem[{Hossain et~al.(2023)Hossain, Vafaei, Azar, Rahman, Farahmandi, and Tehranipoor}]{hossain2023socfuzzer}
Hossain, M.~M.; Vafaei, A.; Azar, K.~Z.; Rahman, F.; Farahmandi, F.; and Tehranipoor, M. 2023.
\newblock Socfuzzer: Soc vulnerability detection using cost function enabled fuzz testing.
\newblock In \emph{2023 Design, Automation \& Test in Europe Conference \& Exhibition (DATE)}, 1--6. IEEE.

\bibitem[{Karri et~al.(2010)Karri, Rajendran, Rosenfeld, and Tehranipoor}]{karri2010trustworthy}
Karri, R.; Rajendran, J.; Rosenfeld, K.; and Tehranipoor, M. 2010.
\newblock Trustworthy hardware: Identifying and classifying hardware trojans.
\newblock \emph{Computer}, 43(10): 39--46.

\bibitem[{Mashnoor et~al.(2025)Mashnoor, Akyash, Kamali, and Azar}]{mashnoor2025llm}
Mashnoor, N.; Akyash, M.; Kamali, H.; and Azar, K. 2025.
\newblock LLM-IFT: LLM-Powered Information Flow Tracking for Secure Hardware.
\newblock In \emph{2025 IEEE 43rd VLSI Test Symposium (VTS)}, 1--5. IEEE.

\bibitem[{OpenAI(2024)}]{openai2024gpt4o}
OpenAI. 2024.
\newblock Hello GPT-4o.
\newblock \url{https://openai.com/index/hello-gpt-4o/}.
\newblock Accessed: 2025-07-11.

\bibitem[{Orenes-Vera et~al.(2021)Orenes-Vera, Manocha, Wentzlaff, and Martonosi}]{orenes2021autosva}
Orenes-Vera, M.; Manocha, A.; Wentzlaff, D.; and Martonosi, M. 2021.
\newblock Autosva: Democratizing formal verification of rtl module interactions.
\newblock In \emph{2021 58th ACM/IEEE Design Automation Conference (DAC)}, 535--540. IEEE.

\bibitem[{Orenes-Vera, Martonosi, and Wentzlaff(2023)}]{orenes2023using}
Orenes-Vera, M.; Martonosi, M.; and Wentzlaff, D. 2023.
\newblock Using llms to facilitate formal verification of rtl.
\newblock \emph{arXiv preprint arXiv:2309.09437}.

\bibitem[{Pearce et~al.(2025)Pearce, Ahmad, Tan, Dolan-Gavitt, and Karri}]{pearce2025asleep}
Pearce, H.; Ahmad, B.; Tan, B.; Dolan-Gavitt, B.; and Karri, R. 2025.
\newblock Asleep at the keyboard? assessing the security of github copilot’s code contributions.
\newblock \emph{Communications of the ACM}, 68(2): 96--105.

\bibitem[{Rajendran et~al.(2023)Rajendran, Tarek, Hicks, Kamali, Farahmandi, and Tehranipoor}]{rajendran2023hunter}
Rajendran, S.~R.; Tarek, S.; Hicks, B.~M.; Kamali, H.~M.; Farahmandi, F.; and Tehranipoor, M. 2023.
\newblock Hunter: Hardware underneath trigger for exploiting soc-level vulnerabilities.
\newblock In \emph{2023 Design, Automation \& Test in Europe Conference \& Exhibition (DATE)}, 1--6. IEEE.

\bibitem[{Saha et~al.(2024)Saha, Yahyaei, Saha, Tehranipoor, and Farahmandi}]{saha2024empowering}
Saha, D.; Yahyaei, K.; Saha, S.~K.; Tehranipoor, M.; and Farahmandi, F. 2024.
\newblock Empowering hardware security with llm: The development of a vulnerable hardware database.
\newblock In \emph{2024 IEEE International Symposium on Hardware Oriented Security and Trust (HOST)}, 233--243. IEEE.

\bibitem[{Solt, Gras, and Razavi(2022)}]{solt2022cellift}
Solt, F.; Gras, B.; and Razavi, K. 2022.
\newblock $\{$CellIFT$\}$: Leveraging cells for scalable and precise dynamic information flow tracking in $\{$RTL$\}$.
\newblock In \emph{31st USENIX Security Symposium (USENIX Security 22)}, 2549--2566.

\bibitem[{Synopsys(2015)}]{spyglass}
Synopsys. 2015.
\newblock SpyGlass: Early Design Analysis Tools for SoCs.
\newblock \url{https://www.synopsys.com/verification/static-and-formal-verification/spyglass.html}.
\newblock Accessed: 2025-07-11.

\bibitem[{Takamaeda-Yamazaki(2015)}]{takamaeda2015pyverilog}
Takamaeda-Yamazaki, S. 2015.
\newblock Pyverilog: A python-based hardware design processing toolkit for verilog hdl.
\newblock In \emph{Applied Reconfigurable Computing: 11th International Symposium, ARC 2015, Bochum, Germany, April 13-17, 2015, Proceedings 11}, 451--460. Springer.

\bibitem[{Webber(2012)}]{10.1145/2384716.2384777}
Webber, J. 2012.
\newblock A programmatic introduction to Neo4j.
\newblock In \emph{Proceedings of the 3rd Annual Conference on Systems, Programming, and Applications: Software for Humanity}, SPLASH '12, 217–218. New York, NY, USA: Association for Computing Machinery.
\newblock ISBN 9781450315630.

\bibitem[{Wu et~al.(2024)Wu, Bansal, Zhang, Wu, Li, Zhu, Jiang, Zhang, Zhang, Liu et~al.}]{wu2024autogen}
Wu, Q.; Bansal, G.; Zhang, J.; Wu, Y.; Li, B.; Zhu, E.; Jiang, L.; Zhang, X.; Zhang, S.; Liu, J.; et~al. 2024.
\newblock Autogen: Enabling next-gen LLM applications via multi-agent conversations.
\newblock In \emph{First Conference on Language Modeling}.

\bibitem[{Yamaguchi et~al.(2014)Yamaguchi, Golde, Arp, and Rieck}]{yamaguchi2014modeling}
Yamaguchi, F.; Golde, N.; Arp, D.; and Rieck, K. 2014.
\newblock Modeling and discovering vulnerabilities with code property graphs.
\newblock In \emph{2014 IEEE symposium on security and privacy}, 590--604. IEEE.

\bibitem[{Yasaei et~al.(2022)Yasaei, Chen, Yu, and Al~Faruque}]{yasaei2022hardware}
Yasaei, R.; Chen, L.; Yu, S.-Y.; and Al~Faruque, M.~A. 2022.
\newblock Hardware trojan detection using graph neural networks.
\newblock \emph{IEEE Transactions on Computer-Aided Design of Integrated Circuits and Systems}, 44(1): 25--38.

\bibitem[{Zhang et~al.(2024{\natexlab{b}})Zhang, Wang, Li, Sun, Zhang, Ma, and Liu}]{zhang2024empirical}
Zhang, J.; Wang, C.; Li, A.; Sun, W.; Zhang, C.; Ma, W.; and Liu, Y. 2024{\natexlab{b}}.
\newblock An empirical study of automated vulnerability localization with large language models.
\newblock \emph{arXiv preprint arXiv:2404.00287}.

\bibitem[{Zhao et~al.(2024)Zhao, Qu, Zhang, Li, Li, and He}]{zhao2024static}
Zhao, Y.; Qu, G.; Zhang, Q.; Li, Y.; Li, Z.; and He, J. 2024.
\newblock Static gate-level information flow for hardware information security with bounded model checking.
\newblock In \emph{2024 IEEE 42nd VLSI Test Symposium (VTS)}, 1--7. IEEE.

\bibitem[{Zhong and Wang(2024)}]{zhong2024can}
Zhong, L.; and Wang, Z. 2024.
\newblock Can llm replace stack overflow? a study on robustness and reliability of large language model code generation.
\newblock In \emph{Proceedings of the AAAI conference on artificial intelligence}, volume~38, 21841--21849.

\end{thebibliography}
\end{document}